\documentclass[aps,prc,10pt,onecolumn,amsmath,amssymb,superscriptaddress]{revtex4-2}
\usepackage{graphicx}
\usepackage{color}
\usepackage{lineno}

\newcommand{\be}{\begin{equation}}
\newcommand{\ee}{\end{equation}}
\newcommand{\beq}{\begin{eqnarray}}
\newcommand{\eeq}{\end{eqnarray}}
\newcommand{\ba}{\begin{array}}
\newcommand{\ea}{\end{array}}

\begin{document}

\title{New Experimental Opportunities in Spin-Physics for Meson Beams}

\vspace{3mm}
\date{\today}

\author{\mbox{William~J.~Briscoe}}
\altaffiliation{Corresponding author: \texttt{briscoe@gwu.edu}}
\affiliation{Institute for Nuclear Studies, Department of Physics, The
        George Washington University, Washington, DC 20052, USA}

\author{\mbox{Igor~Strakovsky}}
\affiliation{Institute for Nuclear Studies, Department of Physics, 
    The George Washington University, Washington, DC 20052, USA}

\noaffiliation

\begin{abstract}
During the past several decades a large quantity of high-quality mesonic photo- and electro-production data have been measured at electromagnetic facilities worldwide. By contrast, meson-beam data for these same final states are mostly outdated, largely of poorer quality, or even non-existent, especially those involving spin asymmetries and polarizations. Thus existing meson beam results provide inadequate input to interpret, analyze, and exploit the potential of the new electromagnetic data. To achieve full benefit of these high-precision electromagnetic data, new high-statistics data from measurements with meson beams, with good angle and energy coverage for a wide range of reactions, are critically needed to advance our knowledge in baryon and meson spectroscopy and other related areas of hadron physics. To address this situation, new, state-of-the-art meson-beam facilities are needed. This presentation summarizes unresolved issues in hadron physics and outlines the opportunities and advances that are possible with such facilities.
\end{abstract}

\maketitle

\section{Introduction}
\label{Sec:Intro}

\subsection{The Need for New Meson Factories}
Our knowledge of the baryon spectrum has advanced rapidly over the past decades~\cite{ParticleDataGroup:2022pth}. The progress has been most significant for non-strange baryons, due to the wealth of new, more precise, measurements made at electron accelerators. The majority of these measurements were performed at Jefferson Lab using the CLAS, GlueX, and Hall~A detectors, at MAMI using the Crystal Ball/TAPS detectors, and at ELSA using the Crystal Barrel/TAPS detectors~\cite{Ireland:2019uwn}. While most earlier progress in baryon spectroscopy was based on analysis of meson-nucleon scattering data, particularly pion-nucleon scattering ($\pi N \to \pi N$ and $\pi N \to \pi\pi N$), photon-nucleon interactions offer the possibility of detecting intermediate states with only small branchings to the $\pi N$ channel. Electron facilities are now producing meson photo- and electro-production data of excellent quality and quantity, much of which is related to spin physics. Current meson-beam data for corresponding hadronic final states are of poor quality or non-existent, and thus are inadequate as input to interpret, analyze, and exploit the full potential of the electromagnetic data. Very few results exist using polarized targets and/or analysis of recoil particle spin. To exploit the high-precision electromagnetic data, high-quality data from measurements with meson beams, with good angle and energy coverage, are needed to advance knowledge in baryon and meson spectroscopy and hadron physics in general. We summarize current unresolved issues in hadron physics and outline the opportunities and advances only possible with new meson-beam facilities.

\subsection{Meson Factories of the Second Half of 20th Century}
There were three historic Meson Factories: the Los Alamos Meson Physics Facility (LAMPF) in Los Alamos, New Mexico; the Swiss Institute for Nuclear Research (SIN) (now the Paul Scherrer Institute (PSI)) in Villigen, Aargau; and TRI University Meson Facility (TRIUMF) in Vancouver, British Columbia. These were designed and built to fulfill the prescription that \textit{Meson factories are medium-energy accelerators ($300 - 1000~\mathrm{MeV}$), capable of producing proton beams of much greater intensity than other types of existing machines}~\cite{Bethe:1964ha} or they were \textit{Accelerators that will produce $500 - 1000~\mathrm{MeV}$ nucleons and mesons in beams thousands of times more intense than existing machines will be able to do experiments never before possible, increase precision of others and reveal processes now unknown}~\cite{Rosen:1966l}.

Two proposals for kaon factories failed to secure funding: KAON at TRIUMF would have provided $100~\mathrm{\mu A}$ of $30~\mathrm{GeV}$ protons (3 MW) to provide four secondary meson beams and a dedicated line for neutrino physics. At LAMPF there was a competing design for the Advanced Hadron Facility (AHF) that would have allowed extraction of a short duty-factor neutrino beam and a high duty factor beam for production of secondary particles.  

BNL successfully operated kaon beam lines at the $30~\mathrm{GeV}$ AGS for many years.Today the  J-PARC facility in Japan is the only Kaon factory in the world. J-PARC is a several-hundred kW proton accelerator with an extensive array of secondary kaon channels. More details about Meson Factories are in Ref.~\cite{Briscoe:2023b}.

\subsection{What Would a Modern Meson Facility be?}
Our recent White Paper~\cite{Briscoe:2015qia} outlined physics programs that could be advanced with new hadron-beam facilities. Studies of baryon spectroscopy, particularly searching for \textit{missing resonances}~\cite{Koniuk:1979vw} with hadronic beam data combined with photo- and electro-production data using coupled-channel analyses. A meson beam facility would also advance the study of strangeness in nuclear and hadronic physics where final-state hyperons are self-analyzing.

A CM energy range up to $2.5~\mathrm{GeV}$ offers  opportunities with pion and kaon beams to study baryon and meson spectroscopy complementary to the electromagnetic programs at \textit{e.g.}, JLab, MAMI, ELSA, SPring-8, and BEPC. This would provide opportunities to contribute to a full evaluation of the high-quality EM data. This is not a competing effort, but rather an experimental program that would 
(i) provide the hadronic complement of ongoing EM programs, and (ii) provide common ground for better, more reliable, phenomenological and theoretical analyses~\cite{Briscoe:2015qia}.

Extensive programs measuring exclusive meson production are planned with the future Electron-Ion Collider (EIC) and J-PARC using charged pion and kaon beams. JLab Hall~D plans a neutral kaon beam facility (KLF) in 2026-2028~\cite{KLF:2020gai}. 


\subsection{Hyperon Spectrum is Important}
The recent PDG2022~\cite{ParticleDataGroup:2022pth} claims about 100 baryon resonances~\cite{ParticleDataGroup:2022pth}. $SU(6)\times O(3)$ implies 400+ baryon resonances if the $3\times 70$-plets and $4\times 56$-plets were completely occupied~\cite{Santopinto:2014opa}. LQCD predicts a similar number of baryon resonances~\cite{Edwards:2012fx}. Using Poincare´s covariant Faddeev equation, one predicts the masses of 90 states not yet seen empirically~\cite{Qin:2019hgk}. The solution of this ``missing'' baryon puzzle~\cite{Koniuk:1979vw} requires more measurements and phenomenological PWAs.
In order to compare directly with QCD-inspired models and Lattice QCD predictions, there has also been a considerable effort to find ``hidden'' or ``missing'' resonances, predicted by quark models and LQCD but not yet confirmed.  Actually, PDG2022~\cite{ParticleDataGroup:2022pth} reports a third of predicted states by CQMs and LQCD.

Recent studies that compare CQM and LQCD results of thermodynamic, statistical Hadron Resonance Gas models, and ratios between measured yields of different hadron species in heavy ion collisions provide indirect evidence for presence of ``missing'' resonances in these contexts (see, for instance, Ref.~\cite{Amaryan:2016ufk}). The contribution order is: hyperons, non-strange baryons, mesons, and light nuclei. It means that hyperons are playing leading role to reproduce Chemical Potential.  In particular, KLF and J-PARC have a link to ion-ion high energy facilities as CERN and BNL and will allow understand formation of our world from quarks and gluons in several microseconds after Big Bang.

\subsection{Spectroscopy of Baryon and Hyperon Resonances: Pion and Kaon Beams}
Most existing meson data on $\pi N$, $\eta N$, $K\Lambda$, $K\Sigma$, \textit{etc.} are 30 -- 50 years old. In many cases, systematic uncertainties were not reported or underestimated. Analyses agree on 4-star resonances (PDG rate~\cite{ParticleDataGroup:2022pth}) visible in elastic $\pi N$ scattering; poor pion-induced data led different analyses to claim different resonance content with no agreement on resonances that couple only weakly to the $\pi N$ channel. The first hyperon $\Lambda(1116)1/2^+$ was discovered back to 1950~\cite{Hopper:1950} while the first determinations of hyperon pole position for $\Lambda(1520)3/2^-$ obtained only recently~\cite{Qiang:2010ve}. 

\begin{figure}[ht]
\vspace{-0.3cm}
\centering
{
    \includegraphics[width=0.7\textwidth,keepaspectratio]{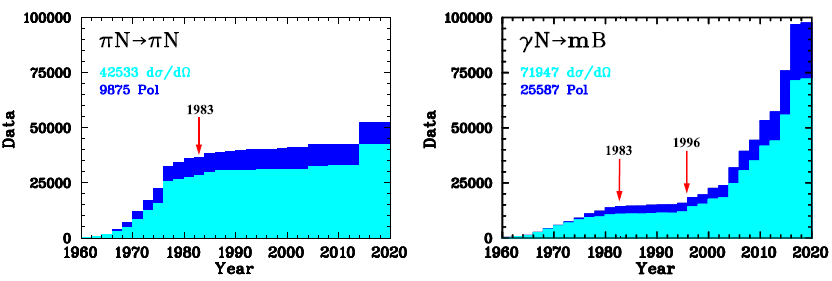} 
}
\centerline{\parbox{1\textwidth}{
\caption[] {\protect\small
\underline{Left}: Stacked histogram of the full experimental pion–nucleon scattering database including 
$\pi^\pm p\to\pi^\pm p$ and $\pi^-p\to\pi^0n$. 
\underline{Right}: Stacked histogram of the full database for single meson photoproduction 
$\gamma N\to mB$, $m = (\pi, \eta, \eta', K, \omega)$, $B = (n, p, \Lambda, \Sigma)$. Light shaded – cross sections,
dark shaded – polarization data. The experimental data is from the SAID database~\cite{Briscoe:2020zzz}.} 
\label{fig:hist1} } }
\end{figure}

As can be seen in Fig.~\ref{fig:hist1}(left), most data for $\pi N$ scattering were obtained before 1983. On the other hand, the '80s saw several advances in accelerator technology that enabled production of photon beams of the order of a GeV in energy; photon energy could be accurately determined through tagging the  electrons degraded via bremsstrahlung, or via laser back scattering. These facilities initially concentrated on photonuclear research, but when threshold for pion production was reached, it became clear that photon beams for hadron physics was a reality; it took a while for this to be realized, which is demonstrated in Fig.~\ref{fig:hist1}(right). This plot shows that a increase in the worldwide data set for photoproduction only occurred after the turn of the 21st century. One can readily see that by 1983, the amount of new dataobtained (both pion- and photoproduction) was tailing off. New intense kaon beams would provide opportunities to locate ``missing'' resonances and establish properties of decay channels for higher excited states.
\section{Studies with Meson Beams}
\subsection{Studies with Pion Beams}
\begin{figure}[ht]
\vspace{-0.3cm}
\centering
{
    \includegraphics[width=0.7\textwidth,keepaspectratio]{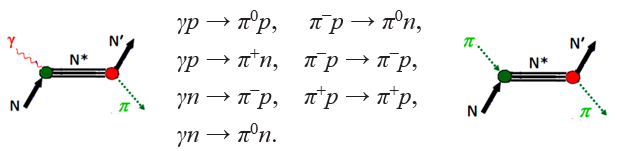} 
}

\centerline{\parbox{0.8\textwidth}{
\caption[] {\protect\small
Photo and pion induced reactions for single pions.} 
\label{fig:pin1} } }
\end{figure}

Measurements of final states involving a single pseudoscalar meson and a spin-1/2 baryon are important. Reactions involving $\pi N$ channels include as shown on Fig.~\ref{fig:pin1}. Data bases for these photo reactions are much larger than for pion reactions (Fig.~\ref{fig:hist1}). The dramatic improvement in statistics made possible in modern experimental physics (EPECUR) is demonstrated (for medium energies).
A similar improvement of the data for low energies is called for.
%
\vspace{-0.5cm}
\subsubsection{Other Studies with Pions}
$\pi N$ elastic scattering data allowed establishment of the 4-star PDG resonances~\cite{ParticleDataGroup:2022pth}. $\pi N\to \pi N$ data are very old, have high systematic uncertainties and are incomplete. Few data $A$ and $R$ for $\pi^-p\to\pi^-p$ and $\pi^+p\to\pi^+p$ and for few energies and angles~\cite{Briscoe:2020zzz}.
No $A$ and $R$ data for $\pi^-p\to\pi^0n$ and very few $P$ data~\cite{Briscoe:2020zzz}; these observables are needed to construct unbiased partial-wave amplitudes. Only by combining information from analyses of both $\pi N$ elastic scattering and $\gamma N\to\pi N$ that it is possible to determine the $A_{1/2}$ and $A_{3/2}$ helicity couplings for $N^\ast$ (see, for instance, Ref.~\cite{Briscoe:2023gmb}).

\begin{figure}[ht]
\vspace{-0.3cm}
\centering
{
    \includegraphics[width=0.4\textwidth,keepaspectratio]{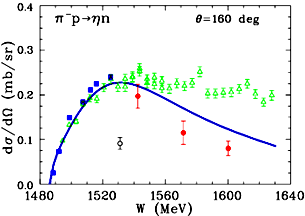} 
}

\centerline{\parbox{1\textwidth}{
\caption[] {\protect\small
Excitation function for the differential cross sections for the reaction $\pi^-p\to\eta n$. Data came from BNL, RHEL, and Saclay. All references are available at Ref.~\cite{Arndt:2005dg}.} 
\label{fig:eta1} } }
\end{figure}

Reactions including $\eta N$ and $K\Lambda$ channels notable – have pure isospin-1/2 contributions: 
\begin{equation}
    \begin{aligned}
    \gamma p\to \eta p \>, ~~ & ~~~~ \pi^-p\to\eta n \>, \\
    \gamma p\to K^+\Lambda \>, ~~ & ~~ \pi^-p\to K^0\Lambda \>, \\
    \gamma n\to \eta n \>,~~&~~~~~\\
    \gamma n\to K^0 \Lambda \>.~~&~~~~~    
    \end{aligned}
    \label{eq:eq4}
\end{equation}
%
Most $\pi^-p\to\eta n$ data published in 1970s unreliable above $W = 1620~\mathrm{MeV}$ (Fig.~\ref{fig:eta1}). Measurements of $\pi^-p\to\eta n$  taken in 1990’s by Crystal Ball Collaboration~\cite{Arndt:2005dg} – extended to peak of $S_{11}$ at $1535~\mathrm{MeV}$. The SAID coupled channel analysis is not able to fit data except for the blue filed circles which came from BNL. Coupled-channel analyses need precise data for $\pi^-p\to\eta n$.
\subsubsection{Double pion production}
\begin{figure}[ht]
\vspace{-0.5cm}
\centering
{
    \includegraphics[width=0.3\textwidth,keepaspectratio]{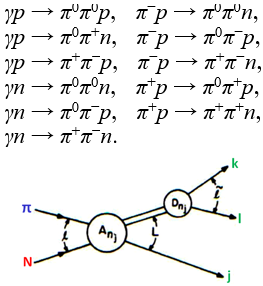}~~~~~~~~~ 
    \includegraphics[width=0.4\textwidth,keepaspectratio]{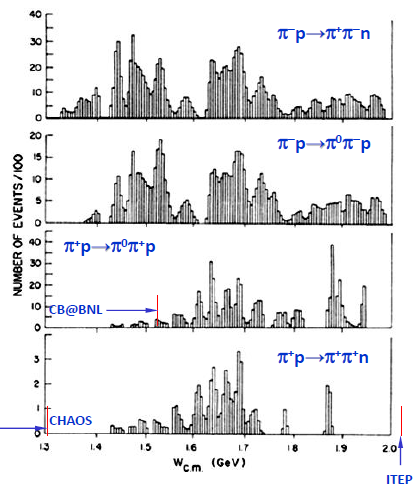} 
}

\centerline{\parbox{1\textwidth}{
\caption[] {\protect\small
\underline{Left}: Double pion production reactions.
\underline{Right}: 241,214 bubble chamber data for double pion production reactions~\cite{Manley:1984jz}.} 
\label{fig:pip1} } }
\end{figure}
Other measurable reactions include $\pi\pi N$ final states (Fig.~\ref{fig:pip1}(left)). Analysis/interpretation more complicated – 
involve 3-body final states. $\pi N \to\pi\pi N$ - lowest energy threshold of inelastic channels and some of the largest cross 
sections. For most established $N^\ast$ and $\Delta^\ast$ resonances, dominant inelastic decays go to $\pi\pi N$ final states. 
Much of $\pi N \to\pi\pi N$ information comes from bubble-chamber data analyzed in isobar-model PWA at $W = 1320$ to 
$1930~\mathrm{MeV}$~\cite{Manley:1984jz} (Fig.~\ref{fig:pip1}(right)).  Since that time the minor contribution for unpolarized measurements came from BNL and TRIUMF and polarized data came from ITEP just for $W = 2060~\mathrm{MeV}$. Need high-quality, high-statistics data for $\pi N \to\pi\pi N$ data that can be analyzed together with complementary data for $\gamma N \to\pi\pi N$ channels. 
%

\subsubsection{Inverse Pion Electroproduction}
Electromagnetic form factors in function  of the four-momentum transfer $k$ from electron  to a hadron, which is actually the four-momentum of the virtual photon, describes  the spatial structure of  hadrons, in particular their mean-square radii. The timelike and spacelike form factors are related by analytic continuation in the complex plane of $k^2$. Accordingly, the complete spatial model of the nucleon must simultaneously describe the nucleon form factors in both regions. Therefore, timelike form factors can be considered as a subject of critical verification in describing the structure of nucleons.

\begin{figure}[ht]
\vspace{-0.5cm}
\centering
{
    \includegraphics[width=0.5\textwidth,keepaspectratio]{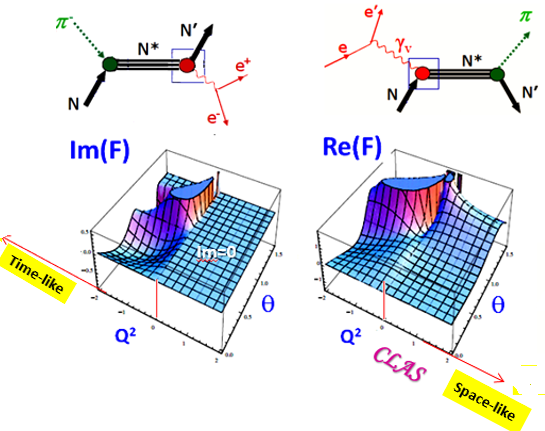} 
}
\centerline{\parbox{0.9\textwidth}{
\caption[] {\protect\small
Inverse pion electroproduction $\pi^-p\to e^+e^- n$ vs pion electroproduction $\gamma^\ast N\to \pi N$.} 
\label{fig:ipe1} } }
\end{figure}

The majority of pion electroproduction data (cross sections and asymmetries) came from CLAS 
(95\%) (\textit{e.g.}, see recent review~\cite{Mokeev:2022xfo}). Inverse Pion Electroproducion (IPE) is the only process, which allows determination of EM nucleon and pion form factors in intervals: $0 < k^2 < 4M^2$ and $0 < k^2 < 4m_{\pi}^2$ which are kinematically unattainable from $e^+e^-$ initial states (Fig.~\ref{fig:ipe1}). $\pi^-p\to e^+e^- n$ measurements will significantly complement current electroproduction. $\gamma^\ast N\to\pi N$ study for the evolution of baryon properties with increasing momentum transfer by investigation of case for time-like virtual photon. Unfortunately, there are no data for IPE~\cite{Baturin:2023b}.
\vspace{0.45cm}

\vspace{-0.45cm}
\subsection{Studies with Kaon Beams}
Related reactions involve the $K\Sigma$ channel (Eqs.~\ref{eq:eq8}). Except for $\pi^+p\to K^+\Sigma^+$, these 
involve mixture of isospin 1/2 and 3/2. There are number of high-quality measurements involving $K \Sigma$ 
photoproduction, status of pion-induced reactions is rather dismal.  Fewer data for $\pi^- p\to K \Sigma$, $\eta' N$, $\omega N$, and $\phi N$ than $\pi^-p$. 

Neutral hyperons $\Lambda^\ast$ and $\Sigma^\ast$ were systematically studied in formation processes as shown in Eqs.~\ref{eq:eq8}. Analyses of $K^-p \to \pi^0\Lambda$ yield fits agree for differential cross section and polarization, but differ for spin rotation parameter $\beta$.  Data for $\beta$ are limited to measurements at only several energies. $\beta$ more sensitive to contributions from high partial waves than differential cross section or polarization. Need new measurements with polarized target. In addition, $\Sigma^{\ast -}$ can be produced in $K^-n$ reactions with a deuteron target:
\begin{equation}
    \begin{aligned}
    K^-p\to K^-p \>, ~~ & ~~ K^-p\to\pi^+\Sigma^- \>, ~~ & ~~ K^-n\to \pi^-\Lambda \>, \\
    K^-p\to K^0n \>, ~~ & ~~ K^-p\to\pi^0\Sigma^0 \>, ~~ & ~~ K^-n\to \pi^0\Sigma^- \>, \\
    K^-p\to \pi^0\Lambda \>, ~~ & ~~ K^-p\to\pi^-\Sigma^+ \>, ~~ & ~~ K^-n\to \pi^-\Sigma^0 \>.    
    \end{aligned}
    \label{eq:eq8}
\end{equation}
\vspace{0.45cm}

PWA is powerful for disentangling overlapping states, especially above $1.6~\mathrm{GeV}$ for $\Lambda^\ast$ ($\Sigma^\ast$) resonances. Let us note that $\Lambda(1405)1/2^-$ and $\Sigma(1385)3/2^+$ lie below the $KN$ threshold; properties of these states obtained only through production processes such as 
\begin{equation}
    K^-p \to \pi^- \Sigma^{\ast +} \to \pi^-\pi^+\Lambda^\ast \>.
    \label{eq:eq8a}
\end{equation}
$t$-channel process provides ``virtual'' $K^0$ beam – allows production of $\Sigma^{\ast +}$. $\Lambda(1405)1/2^-$ and $\Sigma(1385)3/2^+$ identified in decay of $\Sigma\pi$ and $\Lambda\pi/\Sigma\pi$.
%

\subsubsection{K-long Experiment at JLab}
The K-Long project approved to build a secondary KL beam line in Hall~D at Jefferson Lab with a flux three order of magnitude higher than SLAC~\cite{KLF:2020gai}. Scattering experiments on both proton and neutron targets. 
First hadronic facility at Jefferson Lab. It will measure differential cross sections and (self) polarization of hyperons with GlueX detector to enable precise PWA to determine all resonances up to $W = 2500~\mathrm{MeV}$.
KLF experiment will look for ``missing'' hyperons~\cite{Koniuk:1979vw} via reactions, for instance, as shown in Table~\ref{tbl:res1}, 
Additionally, it will perform strange meson spectroscopy to locate pole positions in I = 1/2 and 3/2 channels.
\begin{table}[htb!]
\vspace{-0.3cm}
\centering \protect\caption{The list of reactions which which allows to study some of hyperon resonances.}

\vspace{2mm}
{%
\begin{tabular}{|c|c|}
\hline
Hyperon        & Reaction   \\
\hline
$\Sigma^\ast$  & $K_Lp\to \pi\Sigma^\ast \to \pi\pi\Lambda$ \\
$\Lambda^\ast$ & $K_Lp\to \pi\Lambda^\ast \to \pi\pi\Sigma$ \\
$\Xi^\ast$     & $K_Lp\to K\Xi^\ast \to \pi K\Xi^\ast$ \\
$\Omega^\ast$  & $K_Lp\to K^+K^+\Omega^\ast$ \\
\hline
\end{tabular}} \label{tbl:res1}
\end{table}
%

\begin{figure}[ht]
\vspace{-0.3cm}
\centering
{
    \includegraphics[width=0.5\textwidth,keepaspectratio]{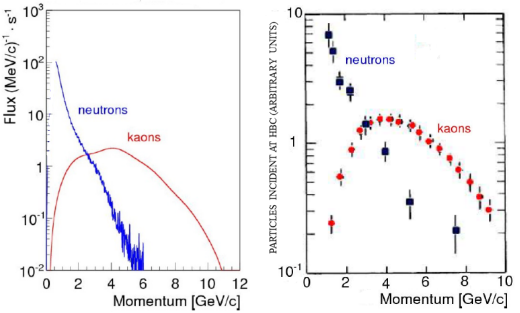} 
}

\centerline{\parbox{1\textwidth}{
\caption[] {\protect\small
\underline{Left}: Rate of $K_L$ (red) and neutrons (blue) on the LH$_2$/LD$_2$ cryogenic target of Hall~D at $12~\mathrm{GeV/c}$ as a function of their generated momenta, with a total rate of $1 \times 10^4~\mathrm{K_L/sec}$ and $6.6 \times 10^5~\mathrm{n/sec}$, respectively. 
\underline{Right}: Experimental data from SLAC measurements using a $16~\mathrm{GeV}$ electron 
beam~\cite{Brandenburg:1972pm}.
} 
\label{fig:yield1} } }
\end{figure}

Hall~D is getting new basic equipment for the KLF project~\cite{KLF:2020gai}: Compact Photon Source (CPS) located in the Tagger Hall, Kaon Production Target (KPT) located in 
the Collimator Cave, and Kaon Flux Monitor (KFM) located in the Experimental 
Hall just in front of the GlueX spectrometer. 
\vspace{0.45cm}
Figure~\ref{fig:yield1} demonstrates that our simulations for the KLF kaon and neutron flux at $12~\mathrm{GeV}$ (left) are
in a reasonable agreement with the $K_L$ and neutron spectra measured by SLAC at $16~\mathrm{GeV}$~\cite{Brandenburg:1972pm} (right).
Let us note that our JLab flux has a factor of 1000 higher than SLAC had in the past.
\vspace{0.45cm}

\begin{figure}[ht]
\vspace{-0.6cm}
\centering
{
    \includegraphics[width=0.4\textwidth,keepaspectratio]{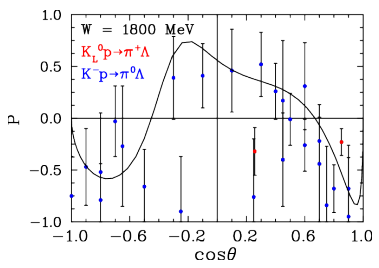}    
}

\vspace{0.2cm}
\centerline{\parbox{1\textwidth}{
\caption[] {\protect\small
Comparison of selected polarization data for $K^-p \to \pi^0\Lambda$ and $K_L p\to \pi^+ \Lambda$ at
$W = 1800~\mathrm{MeV}$~\cite{Zhang:2013cua}.} 
\label{fig:pol1} } }
\end{figure}

The quality of the $K_Lp$ differential cross sections is comparable to that for the $K^-p$ data. It would, therefore, be advantageous to combine the $K_ Lp$ data in a new coupled-channel PWA with available $K^-p$ data. Note that the reactions $K_Lp \to \pi^+ \Sigma^0$ and $K_Lp \to \pi^0 \Sigma^+$ are isospin selective (only I = 1 amplitudes are involved) whereas the reactions $K^-p \to \pi^- \Sigma^+$ and $K^- p \to \pi^+ \Sigma^-$ are not. New measurements with a $K_L$ beam would lead to a better understanding of $\Sigma^\ast$ states and would help constrain the amplitudes for $K^- p$ scattering to $\pi \Sigma$ final states. 
Fig.~\ref{fig:pol1} shows that polarized measurements are tolerable for any PWA solutions. So, new high quality measurements are in order.
\vspace{0.45cm}

\begin{figure}[ht]
\vspace{0.5cm}
\centering
{
    \includegraphics[width=0.45\textwidth,keepaspectratio]{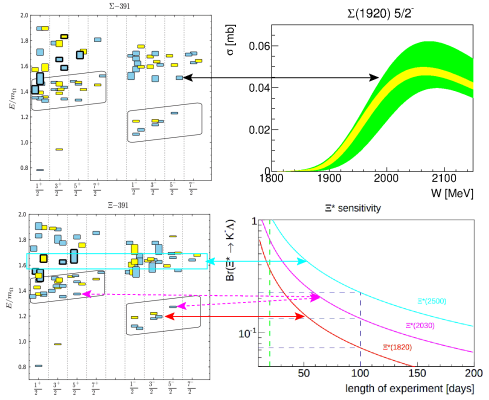} 
}

\centerline{\parbox{1\textwidth}{
\caption[] {\protect\small
Example of comparison between expected KLF measurements for 100 (yellow) and 20 (green) days (right) and LQCD predictions for the hyperon spectrum~\cite{Edwards:2012fx} (left) (for details, see Ref.~\cite{KLF:2020gai}).} 
\label{fig:money1} } }
\end{figure}

As a typical example in Fig.~\ref{fig:money1}, we demonstrate a complete PWA extraction of a fairly low lying but already broad $\Sigma^\ast$ resonance in the reaction $K_Lp \to \Sigma^0 K^+$ assuming 20 (100) days of running.
A clean discrimination of broad excited states
on top of many overlapping resonances with various different quantum numbers is a key feature of
the KLF, unmatched by comparable experiments. The precision of KLF data clearly allows for the
identification of these excited states in a mass range not accessible with previous measurements,
and determination of their quantum numbers and pole positions, which can be then compared with
calculations from LQCD.
\vspace{0.45cm}
\vspace{0.45cm}
\subsubsection{Other Studies with Kaons}
While J-PARC has a whole program of charged strange particle and hypernuclear reactions, the photon beam at KLF 
allows unique access to other channels~\cite{KLF:2020gai}. J-PARC provides separated secondary beam lines up to $2~\mathrm{GeV/c}$. The 
primary beam intensity is currently $50~\mathrm{kW}$, and can be upgraded to $85~\mathrm{kW}$. This will correspond 
to $\sim10^9~\mathrm{ppp}$ (particles per pulse) for pion beam intensity and to $\sim10^6~\mathrm{ppp}$ for negative 
kaon beam flux. 

With $K^-$ beams, currently there is no proposal specific for $S = -1$ hyperons, but the cascades
will be studied in the early stage of E50, hopefully in a few years. The beam momentum bite, $\Delta p/p$, is 
strongly depending on the configuration of the beam line spectrometer, but one can determine beam momentum with the
resolution of $\Delta p/p\sim 10^{-3}$ or $10^{-4}$. 

There is no $K_L$ beamline for hyperon physics at J-PARC. It is 100\% dedicated to the study of
CP-violation. The momentum is spread out from 1 to $4~\mathrm{GeV/c}$.

\section{Summary – What is Needed Today}
The goals of current EM facilities would benefit greatly from having hadron-beam data of a quality similar to that
of electromagnetic data. To this end, it is commonly recognized that a vigorous U.S. program in hadronic physics 
requires a modern facility with pion and kaon beams. In particular, a pion beam and a facility in which $\pi N$ elastic scattering 
and the reactions $\pi^\pm p\to KY$ can be measured in a complete experiment with high precision would be very 
useful. Full solid angle coverage is required to study inelastic reactions such as $\pi^- p\to \eta n$, $\pi^- p\to \phi n$, 
$\pi^\pm N\to \pi\pi N$ or strangeness production (among many other reactions). Such a facility ideally 
should be able to allow baryon spectroscopy measurements up to center-of-mass energies $W$ of about
$2.5~\mathrm{GeV}$, which would require pion beams with momenta up to about $2.85~\mathrm{GeV/c}$. 

Finally, a state-of-the-art hadron beam facility could be used to investigate a much wider range of physics than baryon and meson spectroscopy alone. At the end of the White Paper~\cite{Briscoe:2015qia}, there is a list of endorsers who have expressed support for the initiative described herein: 135 researchers from 77 institutes representing 20 countries around the world.

\section*{Acknowledgments}
This work was supported in part by the U.~S.~Department of Energy, Office of Science, Office of Nuclear Physics, under Awards No.~DE--SC0016583. 



\end{document}